# A Complexity-Informed Approach to Optimise Cyber Defences

Lampis Alevizos
*Volvo Group*
*Amsterdam, The Netherlands*
lampis@redisni.org

***Abstract* — This paper introduces a novel complexity-informed approach to cybersecurity management, addressing the challenges found within complex cyber defences. We adapt and extend the complexity theory to cybersecurity and develop a quantitative framework that empowers decision-makers with strategies to de-complexify defences, identify improvement opportunities, and resolve bottlenecks. Our approach also provides a solid foundation for critical cybersecurity decisions, such as tooling investment or divestment, workforce capacity planning, and optimisation of processes and capabilities. Through a case study, we detail and validate a systematic method for assessing and managing the complexity within cybersecurity defences. The complexity-informed approach based on MITRE ATT&CK, is designed to complement threat-informed defences. Threat-informed methods focus on understanding and countering adversary tactics, while the complexity-informed approach optimises the underlying defence infrastructure, thereby optimising the overall efficiency and effectiveness of cyber defences.**

## I. Introduction

During rapid digital transformation, organizations face an increasingly complex IT landscape. This complexity is further compounded by the layers of cybersecurity measures necessary to protect digital assets. Security becomes a trade-off, and the trade-off is between security and everything else [1]. This trade-off becomes more challenging as both IT systems and security measures grow in complexity. The cybersecurity domain intended to protect and enable business operations, many times becomes an additional layer of complexity on top of an already complex IT landscape. Therefore, this additive complexity oftentimes leads to decreased visibility and observability, increased management overhead, and even potential security gaps [2].

Furthermore, the complexity of the additive cybersecurity layers often hinders clear communication of their value to organizational leadership, creating a significant gap between cybersecurity professionals and board members. The report by MIT and Proofpoint [3] highlights this disconnect, as one of the findings states that 69% of board members believe they see eye-to-eye with their Chief Information Security Officers (CISOs), only 51% of CISOs share this perception. Remarkably, only 67% of board members feel they understand cybersecurity matters well enough to have an informed discussion with their CISO. This communication gap is further evidenced by the fact that just half of board members regularly interact with their CISO, with about a third only seeing the CISO during board presentations. Moreover, the report underlines that CISOs often report on technical security statistics that may not align with the business metrics most relevant to board members. This misalignment in communication and understanding leads to suboptimal resource allocation and a lack of organizational prioritization for cybersecurity initiatives. Therefore, a clear articulation of cybersecurity value and more informed decision-making at the board level becomes imperative. This problem is not limited to the board level. It exists within broader leadership layers of organizations as oftentimes digital transformation, strategy and innovation, platforms and technology or information technology departments casually view cybersecurity as an additional burden rather than a business enabler. Thus, it becomes a pressing need to de-complexify the cybersecurity landscape and produce metrics that can help cybersecurity leaders in their effort to highlight and realize value throughout their organizations. This paper proposes that by simplifying cybersecurity structures and processes, organizations can not only enhance their security posture but also position cybersecurity as a true business enabler.

To address this issue, we adapt and extend the complexity theory model proposed by McNerney et al. [4]. While their work focused on manufacturing processes, we suggest that similar principles can be applied to cybersecurity to reduce complexity and improve efficiency and effectiveness.
This work aims to:

1. Develop a systematic methodology for assessing and reducing complexity in cybersecurity systems.
2. Demonstrate the application of this methodology through a case study using tactics techniques and procedures (TTPs) of the MITRE ATT&CK adversary knowledgebase[1].
3. Propose a framework for translating reduced complexity into clear business value propositions.

We argue that cybersecurity professionals can better manage their domains, make more informed decisions, and ultimately expand their positive influence across the broader IT landscape stakeholders, and thereby their organizations. Lastly, we contribute to both the theoretical understanding of complexity in cybersecurity and provide practical guidance for cybersecurity leaders that intend to enhance the effectiveness and demonstrable value of their security programs.

[1] https://attack.mitre.org/

## II. Literature Review – Complexity in Cybersecurity

The application of complexity theory to cybersecurity has gained significant traction in recent years, offering new perspectives on managing and understanding cybersecurity challenges. The foundational work of McNerney et al. [4] provides the theoretical basis for much of this work. Their model, which relates design complexity (d*) to the rate of performance improvement (α) through the equation $\alpha = 1 / (\gamma d^*)$, has proven applicable beyond its original manufacturing context.

Building on this foundation, Cardenas et al. [5] were among the first to explicitly apply complexity theory to cybersecurity. Their work highlighted how the increasing interconnectedness of cyber-physical systems was leading to emergent behaviours that traditional security models could not address [5]. Although groundbreaking at the time, this early work was primarily theoretical and lacked empirical validation in real-world cybersecurity context. Bartol et al. proposed a framework for measuring information systems security complexity [6]. Their approach is focused on both technical and organizational factors; however, it remains on high-level with broad scope and does not provide specific metrics for individual security controls. We address this limitation by focusing on granular complexity measurements for specific tactics techniques and procedures (TTPs). Tisdale in her work proposed to use systems thinking theory to manage cybersecurity complexity [7]. Although her work is theoretically sound, the approach lacked concrete implementation steps. In our work we build upon the systems thinking approach and provide, through a case study, specific, actionable steps for complexity management.

Recently, a few studies have begun to explore strategies to measure and even reduce cybersecurity complexity. Specifically, Harbertson et al. [8] proposed a framework that assesses complexity in implementing Cybersecurity performance goals through three components: interconnections, organizational capacity, and technical debt. Their work provides for a structured method to quantify complexity using detailed rubrics, demonstrating its application through hypothetical case studies. Although this framework provides a novel way to assess cybersecurity complexity, it requires further empirical validation in real-world scenarios. Moreover, the attributes of interconnections and organizational capacity is driven from high level metrics, as shown in their hypothetical case studies.

Resende et al. [9] conducted a survey on the applications of Kolmogorov complexity in cybersecurity. They proposed a taxonomy categorizing applications into five main domains: human interactions, software, malware, identity/authentication, and theory to practice. Their survey covers applications ranging from phishing detection and text analysis to malware classification and cryptographic protocol validation. While noting the potential of these methods, several limitations arising and areas needing further research, such as real-world validation and optimisation of algorithms for specific cybersecurity contexts.

Yan [10] proposed a systems thinking approach for cybersecurity modelling, highlighting the need to consider factors such as complexity, unpredictability, dynamics, and asymmetry. The paper presents a framework for applying systems thinking to cybersecurity, including stakeholder analysis and a modelling framework that combines mathematical and empirical approaches. Despite Yan's work providing for a valuable holistic perspective that aligns with our complexity-informed approach, it remains largely theoretical and lacks specific application to real-world cybersecurity challenges. Our research builds upon Yan's systems thinking foundation by applying complexity theory more concretely to specific cybersecurity issues, particularly in the context of real-world cyber-attacks utilizing the MITRE ATT&CK knowledge base. Furthermore, where Yan's work provides a broad overview, our study aims to develop quantifiable metrics and practical strategies for managing complexity in operational cybersecurity environments, addressing a gap in the current literature. McEver et al. [11] applied complex adaptive systems theory to develop a framework for resilient cybersecurity. The authors argue that cybersecurity should be viewed as an evolving ecosystem that aims to reduce complexity rather than static defence that adds complexity incrementally. While this approach offers a novel perspective on cybersecurity resilience, it lacks concrete metrics for quantifying complexity in cybersecurity. Our research extends this work by developing quantitative models that can measure the impact of complexity down to specific controls, eventually providing for enhanced decision making.

While these studies have significantly advanced our understanding of complexity in cybersecurity, several important gaps remain. Most notably, there is a lack of empirical validation using real-world data, particularly at the level of specific security controls or techniques. The absence of integration with widely used knowledge bases like MITRE ATT&CK is also a significant limitation, as it hinders the practical application of these theoretical insights.

Furthermore, while many studies have proposed methods for quantifying complexity, there is a shortage of research offering practical, validated strategies for reducing complexity in operational cybersecurity environments. The focus has largely been on system-wide complexity, with limited attention to the granular complexities of individual security measures.

Our study aims to address these gaps by applying McNerney's model to a specific MITRE ATT&CK TTPs using real-world data to empirically validate the relationship between design complexity and security control effectiveness. We develop a practical methodology for assessing and reducing the complexity of specific security controls and proposing a novel version of the Design Structure Matrix (DSM) for cybersecurity, therefore bridging the gap between theoretical models and practical application in this critical field.

## III. Proposed Framework

### A. Adapting complexity theory to cybersecurity

Our study builds upon the complexity theory model proposed by McNerney et al. [4] We adapt it to the specific context of cybersecurity, focusing on identifying and measuring complexity potentially arising from multiple layers of security or controls, placed to defend against MITRE ATT&CK tactics techniques and procedures. This section outlines our theoretical framework and explains how we've modified and extend the original model to pursue our research objectives.

McNerney et al.'s model [4] suggests that the rate of performance improvement (α) for a technology is inversely related to its design complexity (d*) and the inherent difficulty of improving individual components (γ), expressed as:

$$a = 1 \,/\, (\gamma d^{*})$$

In this model, d* is derived from a Design Structure Matrix (DSM) that represents the interdependencies between components of a system. To adapt this model to cybersecurity, we propose the following modifications:

1. **Components**: Instead of physical components, our "components" are the existing security controls in place to defend against a specific TTP.
2. **Design Complexity (d)*:** We redefine d as the interconnectedness of security controls. A higher d* indicates a more complex set of dependencies between different security layers.
3. **Performance Improvement (α)**: In our context, α represents the rate at which the effectiveness of our defences improves against a TTP with either invested effort or tooling over time.
4. **Inherent Difficulty (γ)**: We interpret γ as the inherent challenge in enhancing individual security controls. This may vary based on factors such as technical complexity, e.g., existence of multiple cloud environments, or various operating systems secured by the same controls against a TTP, while the holistic coverage and applicability remains unknown. Resource requirements may be another factor causing intrinsic difficulty, thereby also measured and reflected as γ.

Next, we introduce the concept of the Cybersecurity Design Structure Matrix (CDSM) to map the interdependencies between security controls. The CDSM is a square matrix where:

- Rows and columns represent individual security controls against a TTP.
- Matrix elements indicate whether and how controls interact or depend on each other.

From the CDSM, we construct an algorithm to derive our cybersecurity-specific d*.

### B. Limitations of the original model

The direct application of this model to cybersecurity is limited by the following factors:

1. It does not account for the potential benefits of certain types of complexity in cybersecurity. Primarily, those introduced by the strategy of defence-in-depth. This strategy uses a series of security layers or controls to protect an organization's assets.
2. The model assumes all complexity is detrimental to improvement, which is not always the case in cybersecurity contexts.
3. It does not consider the unique characteristics of cybersecurity, namely, some interdependencies between components (security controls) can enhance the overall cyber defence effectiveness or efficiency.

### C. Defining and Identifying Complexity

Complexity in cybersecurity can be defined as the degree of interdependence and interaction among the various components of an organization's cyber defence. Components may be technological tools, human resources, and operational processes. Complexity is expressed in the number of elements, their interconnections, and the non-linear relationships between them. However, in cybersecurity, complexity can have both positive and negative implications. For instance, a key example of beneficial complexity is the defence-in-depth strategy where multiple layered security controls are used to protect against cyber threats. Nonetheless, excessive complexity leads to increased operational overhead, potential configuration errors, difficulties in incident response, unclear demonstration of cybersecurity value, and challenges in articulating the importance of cybersecurity to stakeholders.

Tesler's Law, also known as the Law of Conservation of Complexity, theorises that every system contains an irreducible level of complexity [12]. Therefore, the objective in cybersecurity is not to eliminate complexity entirely, but to manage it effectively. The goal is to optimise the balance between robust security measures and operational efficiency, thereby positioning cybersecurity as a true business enabler rather than an impediment to business goals, overall growth and innovation.

### D. Novel Enhancements

To address the limitations in section III.C and create a model that is applicable to cybersecurity, we introduce the following key enhancements:

1. **Effective Design Complexity ($\boldsymbol{d_e}$)**: Serves as a new metric that distinguishes between harmful and beneficial complexity and is calculated with the following equation:

$$d_e = d^{*} - \beta(d_b)$$

Where:

- **d*** is the original design complexity derived from our Cybersecurity Design Structure Matrix (CDSM).
- $\boldsymbol{d_b}$ is the measure of beneficial complexity.
- **β** is a weighting factor ($0 \leq \beta \leq 1$) that represents the effectiveness of the beneficial complexity.

2. **Beneficial Complexity ($\boldsymbol{d_b}$)**: We define and quantify beneficial complexity based on three key factors:

$$d_b = \frac{(w1 * Diversity + w2 * Independence + w3 * Coverage)}{(w1 + w2 + w3)}$$

Where:

- **Diversity (0-1)** represents the variety of security controls. A higher score indicates a wider range of distinct control types, thus, enhanced cyber defence.
- **Independence (0-1)** measures the degree to which controls have separate failure modes. A higher score suggests that controls are less likely to fail simultaneously, thus, enhanced cyber resilience.
- **Coverage (0-1)** assesses the range of attack vectors addressed. A higher score indicates broader protection against various threat types.
- **w1, w2, and w3** are context-specific weights.

3. **Enhanced Performance Improvement Model**: Our novel model for the rate of performance improvement (α) becomes:

$$\alpha = \frac{1}{(\gamma(d^* - \beta(d_b)))}$$

4. **Cybersecurity Design Structure Matrix (CDSM)**: Lastly, we introduce the enhanced CDSM that captures the nature of interactions between security controls, rather than dependencies only. Subsequently, interactions can be:

- **Positive:** controls enhance each other's effectiveness.
- **Negative:** controls potentially interfere with each other.
- **Neutral:** controls operate independently.

Our novel framework provides several significant advantages when measuring complexity in cybersecurity. First, we can have a more accurate representation of cyber defences, as we acknowledge and measure the complexity that sometimes may be beneficial. Second, we can now distinguish between complexity that hinders improvement and complexity that enhances security effectiveness and team efficiency. Third, we introduce a quantitative basis for optimising the balance between simplicity, and robust, multi-layered defence. Fourth, we provide cybersecurity experts with a tool to assess whether additional controls will contribute beneficial complexity or merely increase harmful complexity. Lastly, we quantify the impact of complexity on performance improvement, thereby enable cybersecurity executives to have more informed decisions about where to invest resources for maximum security benefit and business value. Thus, fifth, we equip decision-makers with a framework for strategic resource allocation and value realization in cybersecurity investments.

### *E. Research Questions*

We draw the following research questions:

1. **RQ1:** How can we quantify and analyse complexity in cybersecurity while being pragmatic?

2. **RQ2:** How can the quantification of cybersecurity complexity inform decision-making processes in areas such as resource allocation, staffing, and security control optimisation?

3. **RQ3:** What are the potential impacts of distinguishing between harmful and beneficial complexity on the effectiveness and efficiency of cybersecurity systems?

## IV. Case Study

To validate our framework and model for identifying, quantifying, and managing complexity in cybersecurity, we present a detailed case study. The study aims to address RQ1 in a pragmatic manner by applying our model to layers of security controls designed to mitigate specific TTPs of cyber threat actors, as detailed in the MITRE ATT&CK knowledge base.

To ensure practicality and alignment with organizational priorities, we focus our analysis on the security layers mitigating the top 10 TTPs used by adversaries, rather than attempting to quantify complexity across the entire MITRE ATT&CK spectrum. Specifically, we begin with T1059 Command and Scripting Interpreter, identified by MITRE Engenuity as the primary threat vector[2]. As a result, we can conduct a focused and manageable analysis while it also mirrors the prioritization process that organizations should follow in their complexity quantification efforts within cybersecurity.

### *A. Enhanced Cybersecurity Design Structure Matrix*

The first step is to create the enhanced CDSM, which is essential to our complexity analysis. It visually and quantitatively represents the interdependencies between components corresponding to T1059 and our cyber defence. This matrix ultimately allows us to understand how changes in one component might affect others.

We created a square matrix where both rows and columns represent our identified components. We then analysed each pair of components to determine if and how they interact. To effectively and efficiently construct the matrix we:

- Gathered and analysed the technical documentation relevant to T1059 and the mitigating controls.
- Examined grey literature and academic papers on T1059 and the related security controls.
- Interviewed security engineers.
- Scoped and analysed system logs and configuration.
- Conducted controlled tests in a sandbox environment.

[2] https://top-attack-techniques.mitre-engenuity.org/#/top-10-lists

Based on the above process, we scored the interactions using the following scale:

- **X**: Self-interaction
- **0**: Neutral or no significant interaction
- **+1** Positive interaction (components enhance each other's effectiveness)
- **-1**: Negative interaction (components interfere with each other)

Lastly, to normalise bias and guarantee consistency each interaction was independently assessed by two team members with differences being resolved through discussion and, where necessary, additional testing. The full enhanced CDSM is shown in Table 1, and forms the basis for our subsequent complexity calculations.

### *B. Design Complexity (d*) Calculation*

Calculating d* gives us a quantitative measure of our overall complexity introduced by the many security layers used to defend against T1059. This metric helps us understand how interconnected our defences are and identifies potential areas of excessive complexity that might hinder system improvement or management. To calculate d* shown in Table 2, we followed the below steps:

1. Calculated out-degree: Number of components in CDSM it affects (including itself).
2. Calculated in-degree: Number of components in CDSM that affect it (including itself).
3. Determined d_min_i: The minimum of its out-degree and the minimum out-degree of components affecting it.
4. Identified d* as the maximum d_min_i across all components.

*Table 2 – In & out degree of components (TTPs vs Controls).*

| Component | Out-degree | In-degree | Component | Out-degree | In-degree |
|---|---|---|---|---|---|
| **T1059.001 (PowerShell)** | 9 | 9 | **Control 1 (App Whitelist)** | 10 | 10 |
| **T1059.002 (AppleScript)** | 7 | 7 | **Control 2 (Script Monitoring)** | 11 | 11 |
| **T1059.003 (Win CMD)** | 7 | 7 | **Control 3 (PowerShell Constrained)** | 4 | 4 |
| **T1059.004 (Unix Shell)** | 8 | 8 | **Control 4 (Command Logging)** | 11 | 11 |
| **T1059.005 (VB)** | 7 | 7 | **Control 5 (Behaviour Analysis)** | 12 | 12 |
| **T1059.006 (Python)** | 7 | 7 | **Control 6 (Anti-Malware Scan Interface (AMSI))** | 5 | 5 |
| **T1059.007 (JavaScript)** | 7 | 7 | **Control 7 (Net Seg)** | 3 | 3 |
| **T1059.008 (Net Dev CLI)** | 6 | 6 | **Control 8 (Endpoint Detection & Response (EDR))** | 16 | 16 |

The maximum d_min_i is 16 (for control 8, EDR), and therefore d* = 16. This value shows a highly interconnected EDR component. It also suggests that our security controls and the attack techniques they defend against are tightly coupled, which could make our defence, resistant to change and potentially difficult to optimise, specifically when relevant to EDR.

### *C. Beneficial Complexity ($d_b$) Calculation*

Introducing $d_b$ allows us to differentiate between harmful and beneficial complexity. This concept is crucial, because eventually we can differentiate between beneficial complexity (defence-in-depth) and harmful complexity, namely, security layers that act as hindrance rather than business enabler. For each security control, we assessed three factors on a scale of 0 to 1:

1. **Diversity:** Uniqueness of the control in our effort to mitigate T1059.

2. **Independence:** Degree to which the control operates autonomously.

3. **Coverage:** Breadth of T1059 sub-techniques addressed.

These assessments were made based on the existing security controls technical specifications (context-specific), our CDSM, and expert judgment from our cybersecurity team. To mitigate subjectivity, each assessment was independently performed by three team members, with the final score being the average of these assessments. For this case study we used same weight across (w1 = w2 = w3 = 1) but these should be adjusted based on specific organizational context and priorities. The assessment resulted in Table 3.

*Table 3 – Assessing existing security controls against the three key factors for beneficial complexity*.

| Control | Diversity | Independence | Coverage |
|---|---|---|---|
| **Control 1 (App Whitelist)** | 0.8 | 0.6 | 0.9 |
| **Control 2 (Script Mon)** | 0.7 | 0.5 | 1.0 |
| **Control 3 (PS Constrained)** | 0.9 | 0.8 | 0.1 |
| **Control 4 (CMD Logging)** | 0.6 | 0.4 | 1.0 |
| **Control 5 (Behaviour Analysis)** | 0.8 | 0.3 | 1.0 |
| **Control 6 (AMSI)** | 0.9 | 0.7 | 0.4 |
| **Control 7 (Net Seg)** | 0.5 | 0.9 | 0.1 |
| **Control 8 (EDR)** | 0.7 | 0.2 | 1.0 |

Ultimately the $d_b$ calculation is as follows:

$\boldsymbol{d_b}$ = (0.8 + 0.7 + 0.9 + 0.6 + 0.8 + 0.9 + 0.5 + 0.7 + 0.6 + 0.5 + 0.8 + 0.4 + 0.3 + 0.7 + 0.9 + 0.2 + 0.9 + 1.0 + 0.1 + 1.0 + 1.0 + 0.4 + 0.1 + 1.0) / 24 = 16.9 / 24 ≈ 0.7042

Therefore the $d_b$ value of approximately 0.7042 suggests a moderate level of beneficial complexity in our cyber defence against T1059. This indicates that our security controls have a good balance of diversity, independence, and coverage, which contributes positively to our overall cyber defence.

### *D. Effective Design Complexity Calculation ($d_e$)*

To calculate $d_e$ we first need to set β, which is a weighting factor representing the effectiveness of beneficial complexity in offsetting overall complexity. We set this to 0.5, based on our experts' opinions and literature review. Nonetheless, we note that this value will be refined through further research and empirical testing. Therefore:

$$d_e = d^* - \beta(d_b) = 16 - 0.5(0.7042) \approx 15.6479$$

After accounting for beneficial aspects, and while the result is still high, this value is slightly lower than our original d*, indicating that some of our system's complexity contributes positively to its functionality.

### *E. Performance Analysis*

In this step, we quantified the effectiveness of our cyber defence against T1059. We analysed key performance metrics to understand how well our security controls are functioning both individually and as a whole. We gathered data over a 12-month period from security information and event management (SIEM) systems and intrusion detection/prevention systems (IDS/IPS) Organizations should also use other context-specific security telemetry to enrich the dataset.

Performance Metrics:

- **Detection Rate:** Percentage of T1059 successful detection.
- **Prevention Rate:** Percentage of T1059 successful prevention.
- **False Positive Rate:** Percentage of falsely identified T1059 attacks.
- **System Performance Impact:** Percentage increase in system resource usage due to security controls.

Figure 1 shows a steady improvement in detection and prevention rates over time, along with a decrease in false positives. However, we also observe a gradual increase in system performance impact, which is an important finding towards our complexity analysis.

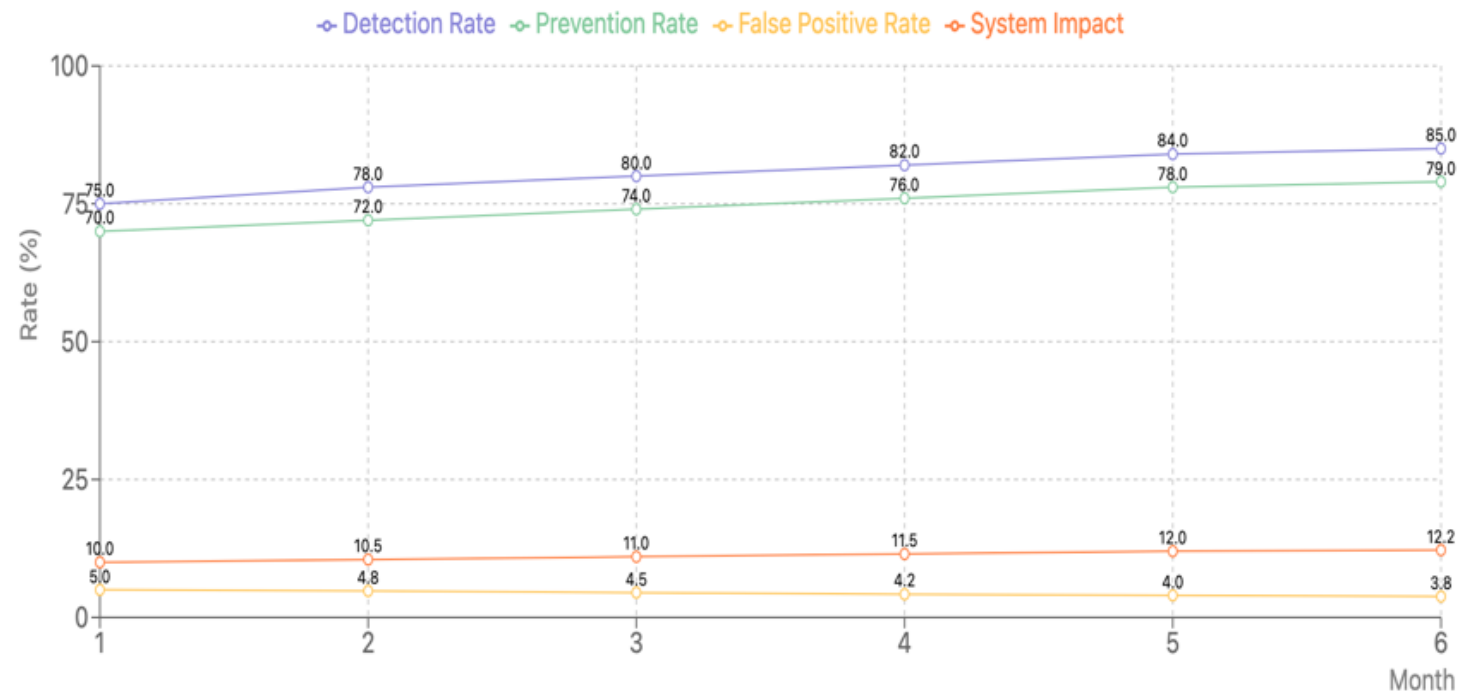

*Figure 1 - Performance Analysis of 4 key metrics.*

### *F. Improvement Rate (a) Calculation*

To calculate the improvement rate (α), we need to plot our performance metrics against time on a log-log scale and calculate the slope using linear regression as shown in Figure 2. This method allows us to quantify how quickly our prevention rate is improving and thereby make predictions about future performance.

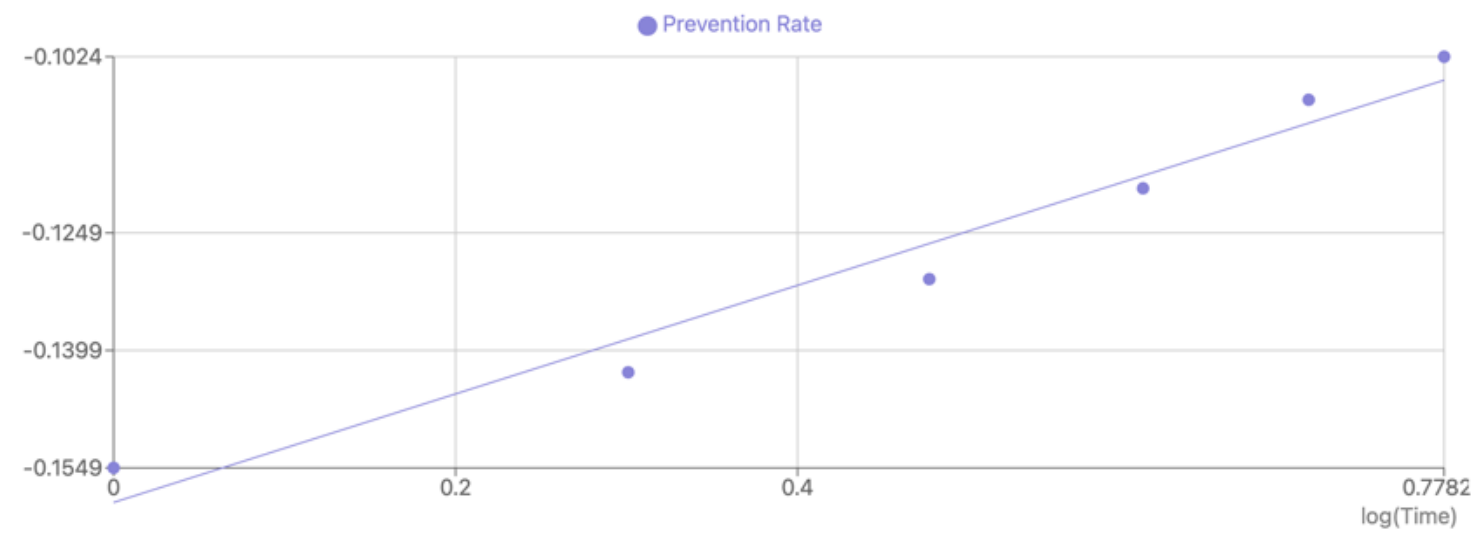

*Figure 2 - Improvement rate (a) calculation.*

This positive value of α indicates that our cyber defence against T1059 is improving over time. The specific value of (0.0757) suggests a moderate rate of improvement, which is typical for complex cybersecurity defences.

### *G. Estimating γ (Difficulty of Improving Components)*

Having the calculations in place, we can now solve for γ:

$$\gamma = \frac{1}{a * d_e} = 1 / (0.0757 * 15.6479) \approx 0.8438$$

γ represents the inherent difficulty of improving our cyber defence against T1059. A value less than 1 suggests that improvements are challenging but achievable.

### *H. Identify Bottlenecks and Opportunities*

Based on the analysis, we identify the following key points:

**Bottlenecks:**

- Control 8 (EDR) has the highest d_min_i (16), which shows a significant bottleneck in our defence.
- Control 5 (Behaviour Analysis) and Control 2 (Script Monitoring) also have high d_min_i values (12 and 11 respectively), indicating that they act as potential bottlenecks.

**Critical components:**

- Control 8 (EDR) affects all other components, making it critical to the overall cyber defence.
- Control 2 (Script Monitoring) and Control 4 (Command-Line Logging) have high coverage across T1059 sub-techniques.

**Opportunities for improvement:**

- Control 7 (Network Segmentation) has the lowest d_min_i (2), thereby it is underutilized in our efforts to defend against T1059.
- Control 3 (PowerShell Constrained Language Mode) has a low d_min_i (4) but high Diversity and Independence scores, showing a good potential for expanded use.

**Performance considerations:**

- Although the detection and prevention rates are improving, the increasing system performance impact shows a need for optimisation.

## V. Discussion

### *A. Complexity-Informed Decision Making*

Organizations can start managing complexity by applying the framework and model immediately. However, it is sensible to create a holistic picture of the complexity introduced by defending against the top 10 TTPs. A complexity heatmap of MITRE ATT&CK could also help visualize the complexities, and therefore facilitate better discussions with various stakeholders within organizations. As an example, we produce a holistic complexity heatmap (Figure 3) for the top 30 TTPs following these steps:

(1) We calculated the effective design complexity $d_e$ for each TTP

(2) Normalize the scores by finding the minimum and maximum $d_e$ values

(3) Apply a simple scaling and colouring scheme, namely, $d_e$ value ranges 0-50 = green = lower complexity, value ranges 51-75 = orange score = moderate complexity, value ranges above 76 = red and represent higher complexity

(4) use python & matplotlib to create the visual.

Applying the model and framework for the top 1 out of the top 10 TTPs, as demonstrated in our T1059 case study, we verify that the complexity metrics provide several improvements to how we achieve cybersecurity that we discuss in this section.

**Targeted Resource Allocation:** Understanding the complexity of our cyber defence allows for a more effective allocation of human and technical resources. As highlighted from the case study, Control 8 (EDR) had the highest d_min_i (16), indicating that it's a critical and complex component. This suggest that we may need to increase capacity or increase the expertise of the personnel to manage and optimise the EDR Conversely, Control 7 (Network Segmentation) had the lowest d_min_i (2), meaning, it is underutilized. This could prompt a re-evaluation of resource allocation or even tooling investment when it comes to defending against T1059.

**Improved Risk Management**: Quantifying complexity means we can anticipate potential issues, thereby allowing for strategic risk mitigation. For instance, the high d* value (16) while defending against T1059 reveals a highly interconnected set of controls. This complexity indicates a higher risk of cascading failures or unintended consequences when making changes either intentionally or by mistake. As a result, risk management strategies can be enhanced to include more thorough change management processes for highly interconnected controls as such and conduct more extensive testing before deploying updates to the EDR system.

**Enhanced Efficiency:** Aligning resources with system complexity means that organizations can optimise their cybersecurity workforce and processes. Our case study shows that while detection and prevention rates against T1059 were improving, system performance impact was increasing. This insight could drive work prioritization to optimise the efficiency of high-complexity controls like EDR and Behaviour Analysis (Control 5). It might also drive work prioritization on process level. For instance, cross collaboration between teams responsible for highly interconnected or depended on controls.

**Focused Capability Development:** Complexity analysis helps identify specific skill sets and technologies required for different areas of our cyber defence. For instance, the high complexity of the EDR system (Control 8) which plays a critical role against T1059, highlights the need for specialized skills in EDR management and optimisation. Therefore, this could be the input for a specific training program or hiring decisions for the team to acquire the necessary expertise. Moreover, the potential for expanding the use of PowerShell Constrained Language Mode (Control 3), which had high Diversity and Independence scores, could drive investment in developing team capabilities in PowerShell security.

**Strategic Course Correction:** Organisation's technological landscape evolves, consequently the cybersecurity complexity may equally follow. Therefore, a security strategy to course-correct accordingly and reflect the changing needs, becomes imperative. Our calculation of improvement rate ($\alpha \approx 0.0757$) and intrinsic difficulty of improvement ($\gamma \approx 0.8438$) can serve as a baseline for monitoring how the cyber defence against T1059 evolves over time. If future measurements show a decrease in α or an increase in γ, it will signal a need to reassess the current strategy. This translates into exploring new technologies to replace or augment existing controls or restructuring the cyber defence to reduce harmful complexity.

**Improved Communication:** Complexity metrics serve as a quantitative basis for explaining cybersecurity needs to organizational leadership, thereby they can facilitate approval for necessary resources. The quantitative approach of our study provides concrete data to support communications with leadership teams. For instance, the high d* value (16) could be used to explain why the security operations centre (SOC) requires significant resources. The moderate improvement rate ($\alpha \approx 0.0757$) could justify additional investment to accelerate improvement, while the $\gamma$ value ($\approx 0.8438$) helps set realistic expectations about the challenges in enhancing the overall cyber defence.

## VI. Conclusions

This paper introduces a novel complexity-informed approach to cybersecurity management, bridging the gap between theoretical models and practical application. We initially adapted and then extended McNerney et al.'s [4] complexity theory to the cybersecurity, developed a quantitative framework that provides insights to decision makers regarding the complexity of cyber defences. The case study against a specific TTP of the MITRE ATT&CK adversary knowledge base demonstrates the practical value of this approach and provides insights on how to achieve balance between beneficial and harmful complexity in cybersecurity defences.

The introduction effective design complexity $d_e$ and beneficial complexity $d_b$ provides cybersecurity leaders with powerful tools for decision-making. These metrics, combined with our proposed cybersecurity design structure matrix, provide for a pragmatic, quantitative approach to manage cybersecurity complexity and apply informed strategies for resource allocation, risk management, and cyber defence optimisation. Lastly, our approach distinguishes that not all complexity is harmful, acknowledging the value of defence-in-depth strategies while identifying areas where simplification can enhance the cyber defence overall.

## VII. Future Research

Several future research directions arise in complexity-informed cybersecurity management that may address current limitations and expand the model's applicability. For instance, developing automated, AI-driven tools for real-time complexity assessment could enhance scalability and adoption. Quantifying the return on investment for complexity reduction efforts can be essential for justifying initiatives to stakeholders. Moreover, adapting the model for distributed infrastructures will address complexity challenges in modern, decentralized IT environments. Another promising future research direction could be the use of machine learning models to predict the impact of complexity changes on system performance, thus enabling proactive complexity management. Lastly, exploring the human factors in cybersecurity complexity can lead to strategies that simplify systems while at the same time enhance human performance in managing these systems. Ultimately, the goal of future research should be to refine and expand the application of complexity theory in cybersecurity, thus, building more resilient and efficient cyber defence across organizations.

*Table 1 – The Enhanced Cybersecurity Design Structure Matrix for T1059 and our mitigating controls.*

| | T1059 .001 | T1059 .002 | T1059 .003 | T1059 .004 | T1059 .005 | T1059 .006 | T1059 .007 | T1059 .008 | CONTROL 1 | CONTROL 2 | CONTROL 3 | CONTROL 4 | CONTROL 5 | CONTROL 6 | CONTROL 7 | CONTROL 8 |
|---|---|---|---|---|---|---|---|---|---|---|---|---|---|---|---|---|
| **T1059.001 (POWERSHELL)** | X | 0 | +1 | 0 | 0 | 0 | 0 | 0 | +1 | +1 | +1 | +1 | +1 | +1 | 0 | +1 |
| **T1059.002 (APPLESCRIPT)** | 0 | X | 0 | +1 | 0 | 0 | 0 | 0 | +1 | +1 | 0 | +1 | +1 | 0 | 0 | +1 |
| **T1059.003 (WIN CMD)** | +1 | 0 | X | 0 | 0 | 0 | 0 | 0 | +1 | +1 | 0 | +1 | +1 | 0 | 0 | +1 |
| **T1059.004 (UNIX SHELL)** | 0 | +1 | 0 | X | 0 | +1 | 0 | 0 | +1 | +1 | 0 | +1 | +1 | 0 | 0 | +1 |
| **T1059.005 (VB)** | 0 | 0 | 0 | 0 | X | 0 | 0 | 0 | +1 | +1 | 0 | +1 | +1 | +1 | 0 | +1 |
| **T1059.006 (PYTHON)** | 0 | 0 | 0 | +1 | 0 | X | 0 | 0 | +1 | +1 | 0 | +1 | +1 | 0 | 0 | +1 |
| **T1059.007 (JAVASCRIPT)** | 0 | 0 | 0 | 0 | 0 | 0 | X | 0 | +1 | +1 | 0 | +1 | +1 | +1 | 0 | +1 |
| **T1059.008 (NET DEV CLI)** | 0 | 0 | 0 | 0 | 0 | 0 | 0 | X | 0 | +1 | 0 | +1 | +1 | 0 | +1 | +1 |
| **CONTROL 1 (APP WHITELIST)** | +1 | +1 | +1 | +1 | +1 | +1 | +1 | 0 | X | 0 | 0 | 0 | +1 | 0 | 0 | +1 |
| **CONTROL 2 (SCRIPT MON)** | +1 | +1 | +1 | +1 | +1 | +1 | +1 | +1 | 0 | X | 0 | +1 | +1 | 0 | 0 | +1 |
| **CONTROL 3 (PS CONSTRAINED)** | +1 | 0 | 0 | 0 | 0 | 0 | 0 | 0 | 0 | 0 | X | 0 | 0 | +1 | 0 | +1 |
| **CONTROL 4 (CMD LOGGING)** | +1 | +1 | +1 | +1 | +1 | +1 | +1 | +1 | 0 | +1 | 0 | X | +1 | 0 | 0 | +1 |
| **CONTROL 5 (BEHAVIOR ANALYSIS)** | +1 | +1 | +1 | +1 | +1 | +1 | +1 | +1 | +1 | +1 | 0 | +1 | X | 0 | 0 | +1 |
| **CONTROL 6 (AMSI)** | +1 | 0 | 0 | 0 | +1 | 0 | +1 | 0 | 0 | 0 | +1 | 0 | 0 | X | 0 | +1 |
| **CONTROL 7 (NET SEG)** | 0 | 0 | 0 | 0 | 0 | 0 | 0 | +1 | 0 | 0 | 0 | 0 | 0 | 0 | X | +1 |
| **CONTROL 8 (EDR)** | +1 | +1 | +1 | +1 | +1 | +1 | +1 | +1 | +1 | +1 | +1 | +1 | +1 | +1 | +1 | X |

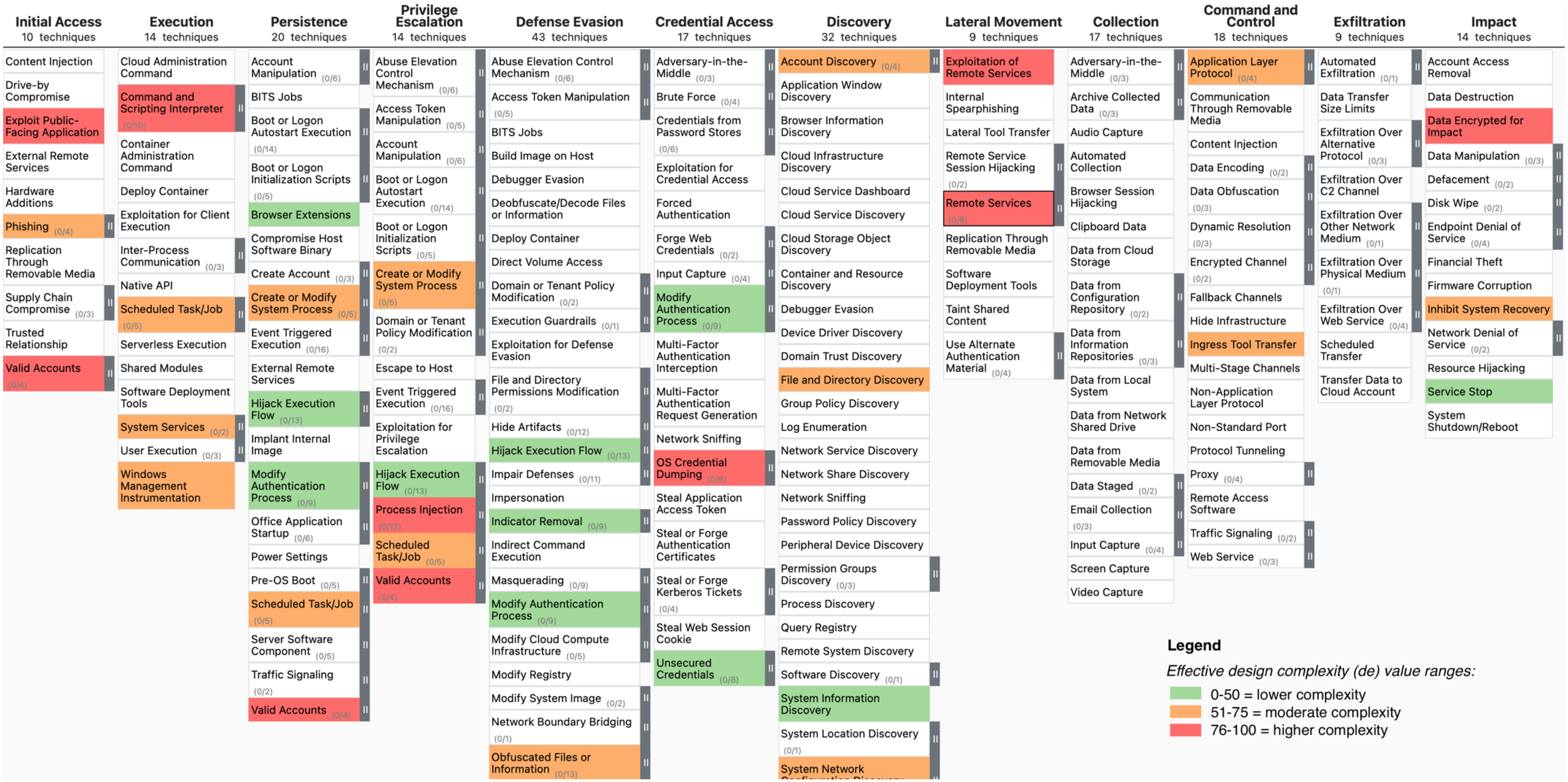

*Figure 3 - Holistic complexity heatmap based on top 30 TTPs.*